\documentclass{aa}
\input psfig.sty
 
\newcommand{\be}{\begin{equation} }
\newcommand{\ee}{\end{equation} }

\newcommand{\apj}{ApJ }

\newcommand{\ltapprox}{\raisebox{-0.5ex}{$\,\stackrel{<}{\scriptstyle\sim}\,$}}
\newcommand{\gtapprox}{\raisebox{-0.5ex}{$\,\stackrel{>}{\scriptstyle\sim}\,$}}
 
\begin{document}
 
\title {VLT Spectroscopy of Galaxies Lensed by Abell AC114:}
\subtitle{Implication for
the Mass Model and the Study of Low-Luminosity Galaxies at High-Redshift
\thanks{Based on observations collected with the ESO Very Large Telescope Antu (UT1) 
and NTT 3.5m Telescope,
and the Hubble Space Telescope.}
}
 
\author{L.E. Campusano\inst{1}
\and  R. Pell\'o\inst{2}
\and J.-P. Kneib\inst{2}
\and J.-F. Le Borgne\inst{2}
\and  B. Fort\inst{3}
\and  R. Ellis\inst{4}
\and  Y. Mellier\inst{3,5}
\and  I. Smail\inst{6}
 }
\offprints{L. Campusano
              \email{lcampusa@das.uchile.cl}}
\institute{Observatorio Cerro Cal\'an, Dept. de Astronom\' \i a,
U. de Chile, Casilla 36-D, Santiago, Chile
\and
Observatoire Midi-Pyr\'en\'ees, UMR 5572,
14 Avenue E. Belin, F-31400 Toulouse, France
\and
Institut d'Astrophysique de Paris, 98 bis boulevard Arago, 75014 Paris,
France
\and
Astronomy 105-24, Caltech, Pasadena CA 91125, USA 
\and
Observatoire de Paris, DEMIRM, 61 avenue de l'Observatorire,
75014 Paris, France
\and
Department of Physics, University of Durham, 
South Road, Durham DH1 3LE, England
}
\date{Received, 2001,  Accepted , 2001}
\authorrunning{L. Campusano et al}
\titlerunning{VLT Spectroscopy of Galaxies Lensed by Abell AC114}
\abstract{
  We present the first results of a spectroscopic survey of faint lensed galaxies
in the core of the galaxy cluster AC114 ($z$=0.312) obtained from observations with the FORS1
spectrograph mounted on the VLT-Antu (Unit Telescope 1). The galaxies were chosen
in areas close to the high-$z$ critical lines predicted by the
gravitational lens model of Natarajan et al(NKSE, 1998) for this cluster,
according to both lensing and photometric redshift criteria.
All the target galaxies are found to correspond to
  background galaxies with redshifts values in the [0.7, 3.5]
  interval. Our spectroscopic observations confirm the predicted
  lensing redshifts for 3 of the multiple-image galaxies, and together
with predictions of the NKSE model led to the discovery
 of a new 5-image configuration at redshift $z=3.347$. A revised NKS model,
compatible with the redshift of this new multiple-image system,
was generated and employed to calculate the gravitational amplifications
of all the observed galaxies. The galaxies corresponding to the
multiple-image systems are found to be intrinsically fainter,
between 0.5 and 1.5 magnitudes,
than the limiting magnitudes of existing {\it blank field} studies.
When all the observed background galaxies are considered, the resulting
intrinsic absolute magnitudes range from
$M_B \sim$ $-$22 to $-$19, with a median value of $-$20.5.
Therefore, a large gain in sensitivity towards low luminosity high-z objects
can actually be obtained, in agreement
with theoretical expectations. This method can be used advantageously
to probe the high redshift Universe and, in particular, its
application to an ensemble of massive cluster cores could
constraint the faint end of luminosity function of high redshift galaxies.
\keywords{Cosmology: observations--- Galaxies: clusters: individual(AC114)--
Galaxies: fundamental parameters--- gravitational lensing---
Galaxies: redshifts -- general -- Methods: data analysis --
spectroscopy}
}
\maketitle
\section{Introduction}
The investigation of the properties of distant galaxies can be
conducted by observing them either in {\it blank fields} or through
the cores of massive clusters of galaxies.  Clusters lenses are
{\it natural telescopes} gravitationally magnifying the background
galaxies, thus allowing the detection of intrinsically fainter and
more distant galaxies than otherwise possible.  Of particular
interest, is the study of spectroscopically confirmed high-$z$
galaxies in cluster cores where the amplification is the highest,
attaining typically values from 1 to 3 magnitudes.
A sample of these galaxies would extend to
fainter magnitudes than the ones currently available from surveys of
the field (Steidel et al 1996a, 1996b, 1998, 1999), and also complement the
latter, because it would be less biased towards intrinsically brighter galaxies.
Therefore, this approach can be used with great advantage to find
still higher redshift galaxies  and to probe the
faint end of the luminosity function at high redshift. 

\begin{figure*}
\hfill
\label{fig1}
\caption{
  {\it HST}/{\it WFPC2} image (F702W) of AC114 cluster core with the
  identified high redshift galaxies and multiple images. The thin
  contour lines represent the total mass distribution as modelled by
  Natarajan et al (1998). The critical lines (dotted lines) are shown
  for $z=3.35$. Circles show the position of the observed E system, and
  crosses correspond to the positions predicted by the NKSE model for the
  counter images of E1.}
\end{figure*}

{\it Natural telescopes} have been successfully used in studying
distant galaxies at almost all wavebands from UV (B\'ezecourt et al
1999), Optical ({\it e.g.} Ebbels et al 1998; Pell\'o et al 1999a, 1999b),
Mid-Infrared (Altieri et al 1999, Metcalfe et al 1999) to Submm 
(Smail et al 1997, Ivison et al 2000). A
difficulty of the optical/near-infrared, much less critical in other
wavebands, is the contamination by cluster galaxies which are not
always easily separated from the background population.  Furthermore,
because of the shallow slope of the galaxy number counts at faint
magnitude, the magnification effect dilutes the density of background
galaxies in the visible/near-infrared bands, hence for a given
magnitude limit, there will be effectively less galaxies seen through
a cluster lens than in {\it blank fields} (Broadhurst 1995, Fort et
al.  1997, Mayen \& Soucail 2000). For the study the luminosity
function of high redshift galaxies, a large number (typically $\sim
100$) of these galaxies are needed. Such number of galaxies, can be
collected through observations, like the ones reported here, of a
sample of about 10 massive cluster-lenses. Our selection criteria of the
distant galaxy candidates are based on a combination of lensing (Kneib
et al  1994, 1996; Ebbels et al  1998) and photometric redshift
criteria (e.g. Mobasher et al 1996,
Lanzetta et al 1996, Gwyn \& Hartwick 1996, Sawicki et al 1997,
Giallongo et al 1998, Fern\'andez-Soto et al 1999, Arnouts et
al. 1999, Furusawa et al 2000, Bolzonella et al 2000). 
Although the lensing criteria in the cluster core is
sufficient (once the mass distribution of the cluster is well
determined) for the selection of distant galaxies (Ebbels et al 1998), 
photometric
redshifts are very effective to select them in the outer part of the
cluster and are also essential in identifying distant galaxies not
resolved by {\it HST} and located in the cluster core. Furthermore,
the photometric properties of the faint galaxies helps to optimise the
instrument choice for the spectroscopic follow-up (visible vs. near-IR
bands).

In this paper, we focus on a particularly interesting cluster-lens,
AC114 (also named ACO S1077, Abell et al 1989), a $z$=0.312 rich
cluster showing a large number of multiple-images at high-$z$ (Smail
et al 1995; Natarajan et al 1998, hereafter NKSE). The redshift of
the gravitational pair S1/S2, $z=1.86$, was obtained by Smail et
al. (1995) with AAT, and later confirmed with NTT with a total
exposure time of $\sim 10$ hours.  Based on this redshift, the
detailed mass-model derived by NKSE predicts the redshifts of 4
multiple images ($A, B, C$ and $D$), ranging from $z\sim$ 1 to 2.5.
Here, we present the results obtained from a spectroscopic survey of
lensed galaxies in AC114 conducted with VLT/FORS1 at Paranal (program
64.O-0439A).  Sections 2 and 3
present a summary of the observations, selection criteria and data
reduction. A detailed discussion of the spectroscopic results is given
in Section 4.  In Section 5, we briefly discuss the improved lens
model for AC114, which now includes the spectroscopic redshift of 2
multiple images as constraints. The properties of these amplified
galaxies are described in Section 6.  Discussion and conclusions are
given in Sections 7 and 8.  Throughout this paper, we adopt $H_0 = 50$
km s$^{-1}$ Mpc$^{-1}$, $\Omega_m =0.3$ and $\Omega_{\Lambda} = 0.7$.

\section{Imaging data and target selection}

All the publicly available images of the cluster core of AC114, to
October 1999, were employed for the selection of background
galaxies. These data include deep {\it HST} images in 3 filter bands:
F555W/{\it WFPC} ($V_{555}$), F702W/{\it WFPC}2 (R),and F814W/{\it
WFPC} ($I_{814}$), as well as U (CTIO) and V (ESO-NTT) images. In
addition, we used deep imaging from a related program with ESO-NTT and
the SOFI camera, including both J and K bands. Unfortunatedly, a
photometric run with SUSI, aimed at obtaining B photometry,
failed because of weather. After our VLT spectroscopic run, we got an existing
deep B-image (courtesy of W. J. Couch), which is taken into
consideration for the discussion about the photometric and lensing
redshifts of the spectroscopically observed galaxies. The basic
characteristics of these images are listed in Table \ref{tab1}, 
including magnitude limits, and references.

\begin{table}
\label{tab1}
\caption[]{ \label{tab1} Characteristics of the images and detection levels.
$3 \sigma$  magnitudes correspond to objects with 4 connected pixels,
each 3 $\sigma$ above the sky level. Observations were carried out at
the {\it HST} (1), ESO NTT (2), the CTIO 4m telescope (3) and the
the AAT (4). }
\begin{flushleft}
\begin{tabular}{lllrrrrrll}
\hline\noalign{\smallskip}
  & $t_{exp}$ & $\sigma$ & pix
& $\lambda_{eff}$ & $\Delta\lambda$ & $ m $ & Ref. \\
  &(ksec)&(\arcsec)&(\arcsec)&(nm)&(nm)&
$3 \sigma$ &   \\
\noalign{\smallskip}
\hline\noalign{\smallskip}
U      & 20.0 & 1.3 & 0.36 & 365 &  40 & 27.1 & a, 3 \\
B      & 9.0 & 1.2 & 0.39 & 443 & 69 & 27.7 & 4 \\
V      & 21.6 & 1.1  & 0.47 & 547 &  53 & 26.7 & b, 2 \\
$V_{555}$& 20.7 & 0.3 & 0.10 & 545 & 105 & 25.8 & b, 1 \\
R      & 16.8 & 0.13 & 0.10 & 694 & 122 & 25.7 &  c,1 \\
$I_{814}$& 20.7 & 0.3 & 0.10 & 801 & 134 & 24.9 & b, 1 \\
J      & 7.2 & 0.9 & 0.29 & 1253 & 169 & 21.5 &  2 \\
K'     & 10.8 & 0.8 & 0.29 & 2164 & 164 & 20.0 &  2 \\
\noalign{\smallskip}
\hline
\end{tabular}
\begin{tabular}{ll}
a) Barger et al 1996 \\
b) Smail et al 1991 \\
c) Natarajan et al 1998 \\
\end{tabular}
\end{flushleft}
\end{table}

In the very central part of the cluster, the high-redshift galaxy
candidates were mainly selected using lensing criteria ({\it eg} 
NKSE).  Photometric redshifts were primarily used for the selection of
background galaxies lying on the outer part of the cluster core, where
lensing criteria are much less efficient, as well as in the core for
cases when lensing failed to give any prediction - specially when the
image of the galaxy was barely resolved. The photometric catalogue was
constructed, after matching the seeing values on the different images, 
using the SExtractor package (Bertin \& Arnouts 1996). Magnitudes in 
this paper refer to the Vega system. Photometric redshifts 
(hereafter $z_{phot}$) were computed through a standard minimization 
procedure, using the {\it hyperz} code (Bolzonella et al 2000). 
This procedure uses a template library of spectra mainly derived 
from the new Bruzual \& Charlot evolutionary code (GISSEL98, 
Bruzual \& Charlot 1993). Photometric redshifts were derived from magnitudes 
computed within a 4$\arcsec$ diameter aperture. In the case of highly distorted images,
located close to bright cluster members, such as A, B and C, 
photometry was obtained through special apertures adapted to the shape 
of each object so that the same
physical region was considered for the magnitude calculations in all
the available images. Because of the lack of a B-band image, the error
bars of the photometric redshifts employed in the target selection
were found to be typically $\sigma_z \sim \pm$ 0.2 to 0.3, with
degenerate solutions in some cases. Given that
the CCD images cover fields of different size, the
photometric selection could be done effectively only on the central
{\it HST} field covering $\sim$ 80$\arcsec\times$80$\arcsec$. Twelve galaxy candidates,
with $z \ge 1.0$ and $R < 24$ were identified in this region.

\begin{table*}
 
\begin{flushleft}
\begin{tabular}{cccccccccc}
\hline\noalign{\smallskip}
 Id.  & r.a.(J2000.0)  &  dec.(J2000.0) & U & B & V & R & $I_{814}$ & J  & K'  \\
\noalign{\smallskip}
\hline\noalign{\smallskip}
A1  & 22:58:49.58  & $-$34:47:53.0 & 23.72 & 23.37 & 22.35 & 23.07 & 23.32 & 22.07 & 21.43
 \\
A2  & 22:58:47.77  & $-$34:48:04.  &       &       &       &       &       &       &
 \\
B2  & 22:58:46.80  & $-$34:47:54.8 & 23.66 & 24.02 & 23.28 & 22.48 &       & 22.54 & 22.25
:  \\
B3  & 22:58:46.53  & $-$34:47:57.1 &       &       &       &       &       &       &
 \\
C3  & 22:58:46.11  & $-$34:47:59.2 & 24.68 & 24.37 & 23.59 & 22.44 & 23.66 &   -   &   -
\\
E1  & 22:58:46.67  & $-$34:48:22.4 &  -    & 25.06 & 24.32 & 23.77 & 23.59 &  -    & 22.68
 \\
S2  & 22:58:48.78  & $-$34:47:54.0 & 22.46 & 22.81 & 22.54 & 22.01 & 22.10 & 21.04 & 19.99
 \\
V1  & 22:58:50.59  & $-$34:48:24.9 & 24.27 & 24.77 & 24.69 & 23.94 & 23.23 & 21.59 & 19.80
  \\
V2  & 22:58:50.72  & $-$34:47:54.3 & 23.49 & 24.27 & 23.91 & 23.20 & 22.84 & 21.71 & 21.06
  \\
V3  & 22:58:51.61  & $-$34:47:59.6 & 24.70 & 25.31 & 24.94 & 23.79 & 22.78 & 21.04 & 19.37
  \\
V4  & 22:58:44.39  & $-$34:48:06.7 & 24.06 & 24.42 & 23.97 & 23.06 & 23.28 & 22.45 & 21.03
  \\
V5  & 22:58:45.75  & $-$34:48:15.8 & 23.80 & 24.50 & 23.84 & 22.97 & 22.68 & 21.10 & 19.25
  \\
\hline\noalign{\smallskip}
V6  & 22:58:50.94  & $-$34:47:26.5 &       & 23.56 & 22.58 & 21.77 & 20.95 & 20.03 & 18.73
 \\
V7  & 22:58:45.60  & $-$34:49:03.9 & 22.88 & 22.97 & 22.16 & 20.66 & 20.15 & 19.17 & 17.68
  \\
V8  & 22:58:54.12  & $-$34:48:28.2 &       & 23.71 &       &       &       & 20.37 & 18.92
 \\
V9  & 22:58:56.56  & $-$34:46:58.6 &       & 22.67 &       &       &       &       & 17.00
  \\
V10 & 22:58:56.98  & $-$34:48:45.8 &       & 25.29 &       &       &       &       & 17.72
  \\
V11 & 22:58:57.46  & $-$34:47:06.8 &       & 23.61 &       &       &       &       & 18.39
 \\
V12 & 22:58:38.15  & $-$34:48:25.3 &       & 22.84 &       & 20.58 &       & 19.30 & 17.43
 \\
V13 & 22:58:38.70  & $-$34:50:08.2 &       & 24.24 &       &       &       & 20.67 & 18.69
 \\
V14 & 22:58:38.71  & $-$34:50:08.5 &       & 24.04 &       &       &       & 21.60 & 19.77
  \\
V15 & 22:58:36.21  & $-$34:49:40.5 &       & 25.28 &       &       &       &       & 17.03
 \\
V16 & 22:58:34.59  & $-$34:49:30.1 &       & 23.08 &       &       &       &       & 18.15
  \\
V17 & 22:58:34.80  & $-$34:50:34.1 &       & 26.26 &       &       &       &       & 18.75
  \\
V18 & 22:58:33.65  & $-$34:49:42.2 &       & 23.40 &       &       &       &       &
  \\
\hline\noalign{\smallskip}
\end{tabular}
\end{flushleft}
\caption[]{ \label{tab_phot}
Position and photometry for the 23 observed background galaxies.
The horizontal line separates the core sample,
for which the strong selection criteria apply. When available, the
identifications are given according to Natarajan et al 1998. "V" objects
have numbers increasing with their distance from the cluster center. When
a space is left blank, the object is "out of the field"
in this filter, whereas non-detected objects are given by "-". V-band
photometry corresponds to ground-based data. These are aperture
magnitudes computed within a $4\arcsec$-diameter aperture, after
correction for seeing differences between the images.
}
\end{table*}

Given that both lensing and photometric redshifts were available for
most of the high-z galaxy candidates lying in the cluster core, and
that the spectroscopic set-up allowed the observation of only a few of
them, we proceeded mostly with the candidates with consistent redshift
predictions. The designation of the observed objects is the same as
that of NKSE, when available, otherwise they are labelled "V" with
numbers increasing with their distance from the cluster center.  
The galaxies in the core that were actually observed, are
identified in Fig. \ref{fig1}. In Table \ref{tab_phot}, positions
and UBVRIJK photometry are given for the all the observed background galaxies.

\section{Spectroscopic Observations and data reduction}

The spectra of the high-redshift galaxy candidates were obtained 
on the night of October 5, 1999, with
the ESO VLT Antu (UT1) telescope and its FORS1 spectrograph working on
the multi-object mode. A slit width of 1 $\arcsec$ and
the grism G300V were used, resulting in a wavelength coverage of
$\sim$4000-8600$\AA$, and a resolution of $R=500$.  Three masks,
or slit configurations, were employed for total exposure times of
2h15m, 1h30m and 1h17m, respectively. Each of them spans a $\sim 7
\arcmin$ field, with 19 to 20 slitlets having a fixed 22 $\arcsec$
length. The fact that the slit lenght was comparable to the angular
size of the cluster core, restricted to $\sim$ 5-6 the number of
strongly magnified galaxy candidates that could be observed per mask. 
Multiple images according to the NKSE model were given the first priority,
and the secondary targets to fill the remaining slitlets inside
the $1.2\arcmin \times 1.2\arcmin$
cluster core were chosen from the photometric high-$z$ list of candidates.
In order to improve the efficiency, we kept
some of the faintest candidates from the cluster core in several slit
configurations while changing the brighter targets lying in the outer
parts of the cluster. Once the prioritary targets had determined the position of
the center of the mask and its orientation, the slits outside the core could be used
only to observe galaxy cluster members. The seeing values during the
observations were in the range $\sim 0.6-0.8\arcsec$, except during
the last 1500 seconds of the exposure with the third mask ($1.3
\arcsec$). Spectra of the spectrophotometric standard star Feige 110
were obtained during twilight for calibration. The data reduction was
done using standard {\sc iraf} packages.

\begin{figure}
\label{fig_2dlz}
\caption{
  Two dimensional spectra of background galaxies showing emission lines in the low redshift
domain: $0.33 \le z \le 0.80$. Galaxies are identified as in Table~\ref{tab_phot}.
The arrows indicate the position of [OII]3727\AA
  line. The wavelength interval 4760\AA--6590\AA is
  shown.  }
\end{figure}
 
\begin{figure}
\label{fig_2dhz}
\caption{
  Two dimensional spectra of background galaxies showing emission lines in the high redshift
domain: $0.80 \le z \le 1.21$. Galaxies are identified as in Table~\ref{tab_phot}.
The arrows indicate the position of [OII]
  line at 3727\AA. The wavelength interval 6620\AA--8500\AA is
  shown.  }
\end{figure}

\section{Spectroscopic Results}

Sixty-two galaxy spectra were extracted from the integrations
performed through the 3 mask configurations selected for use with
FORS1.  The inspection of the wavelength and flux calibrated spectra,
revealed that twenty-seven were galaxies belonging
to AC114, twenty-three galaxies lay in the background of the cluster, and 
ten were foreground galaxies (two spectra remained unidentified). In
this paper, we concentrate on the spectroscopically identified
background galaxies and, in particular, on those located in the central
$1.2\arcmin \times 1.2\arcmin$ region of AC114. The results on the cluster
galaxies will be presented in a forthcoming paper.

\begin{figure*}
\psfig{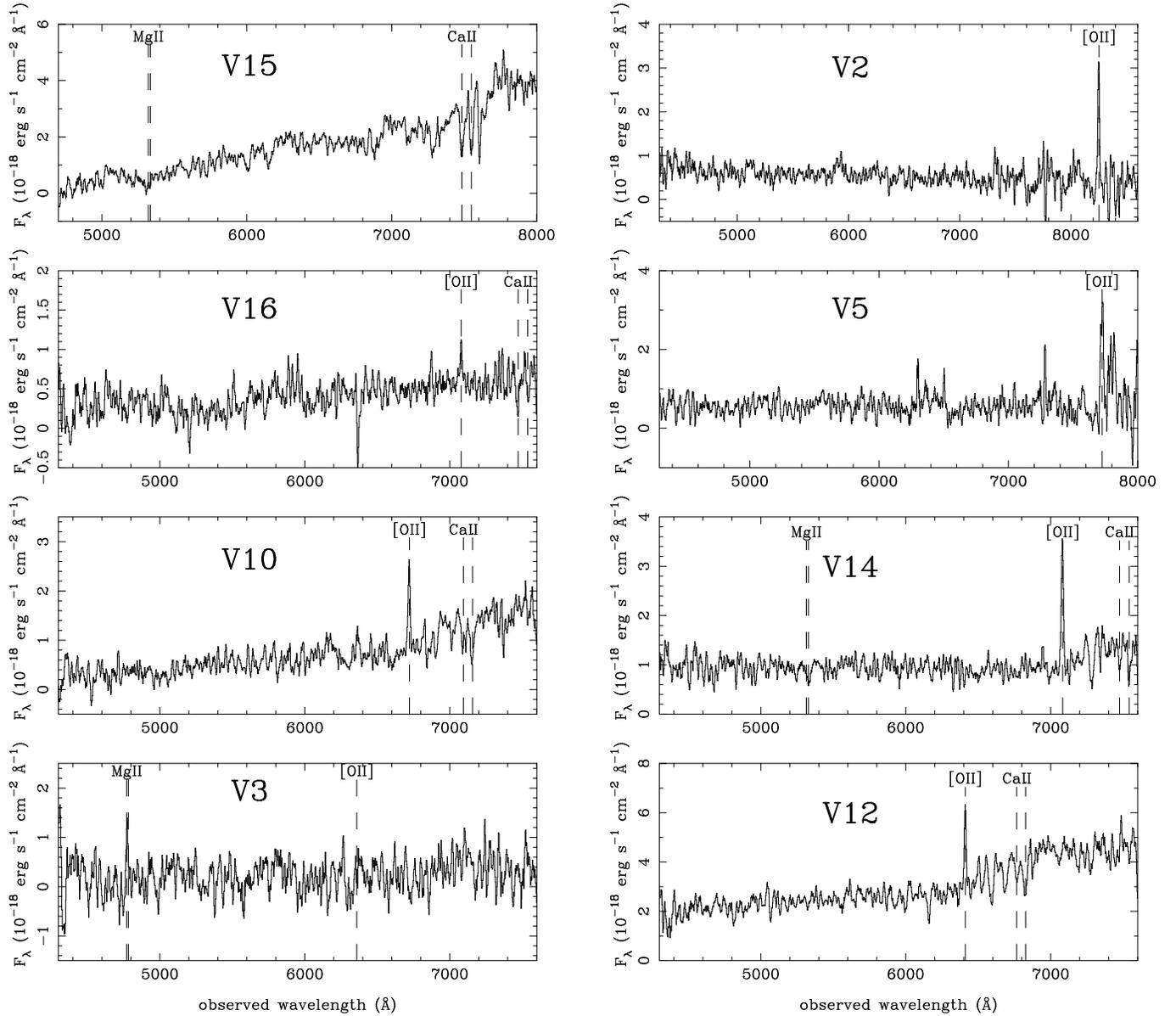}
\label{fig_spec}
\caption{
  Spectra of the identified galaxies in the background of AC114. The label on
  the top of each spectrum gives the identification number, according
  to Table~\ref{tab_phot}, as well as the redshift. Spectra showing
  single emission lines are also displayed as 2D images. Fluxes are
  given in arbitrary units because of the lack of spectrophotometric
  conditions (neither the standard star nor the galaxy are well
  sampled by the slit), but they have the right order of magnitude in
  ergs s$^{-1}$ cm$^{-2}$ \AA$^{-1}$ }
\end{figure*}

The 2-dimensional spectra of the background
galaxies with $0.33 \le z \le 0.80$ are shown in Figure \ref{fig_2dlz},
where the arrows indicate the position of the [OII] line at 3727\AA.
Similarly, Figure \ref{fig_2dhz}, displays the 2-dimensional spectra for
the $0.80 \le z \le 1.21$ interval.
Figure~\ref{fig_spec} displays calibrated one-dimensional
spectra of the observed background
galaxies with $z \gtapprox 1$, showing the identified emission and
absorption features. The designation of the galaxies
is the same as indicated in Table~\ref{tab_phot}.

The redshifts measured for the background galaxies are listed in Table~\ref{tab_z},
together with the identified spectral features either in absorption or emission.
For the objects with a fair signal-to-noise value, the best fit galaxy spectral
type is given, from E to Im.
A discussion on the
redshift determinations of the galaxies lying in the cluster core is
given below, which is specially delicate for the spectra presenting only one strong
spectral feature. Table~\ref{tab_z} summarizes
the redshift determinations and the main spectral features identified.  
Excepting one uncertain case (V1, see Sect.6.5),
all of the 10 high-z candidates selected in the
cluster core region were found to have spectroscopic redshifts between
$\sim 1$ and 3.5,
in good agreement with the redshift selection criteria. The strongly lensed
galaxies in the cluster core are discussed one by one in 
Sect. 6.

For comparison, Table \ref{tab_zphot} lists the lensing (NKSE) redshifts
for the multiple systems and the photometric redshifts for the 
the observed galaxies in the cluster core, all of them with
predicted redshifts greater than $\sim 1$. The only
exception is V3, for which the {\it a posteriori} $z_{phot}$
is lower than the initial estimate, $z \sim 0.8$, and in fair agreement with the
spectroscopic value. Overall, our method to identify lensed galaxies
with $z > 1.0$ had a success rate $\gtapprox 80\%$.
Although the $z_{phot}$ used for target selection had
larger uncertainties than those given in Table \ref{tab_zphot} (because of the lack
of deep B photometry), they were an effective complement to the
$z_{lens}$ specially in the case of barely resolved objects which did
not have a predicted $z_{lens}$, and turned out to be fully compatible with
the spectroscopic redshift within the errors. Table \ref{tab_zphot} also 
lists the $z_{phot}$ for the rest of the observed background galaxies.

>From Table \ref{tab_zphot}, it can be seen that the $z_{lens}$ predicted
by the NKSE for A1/A2 is in remarkable agreement with the measured
redshift for the system. Although the redshift determination of the systems
B and C is more uncertain, they suggest that the $z_{lens}$ predictions
of the NKSE underestimate the actual values.

\begin{table}
\begin{flushleft}
\begin{tabular}{ccl}
\hline\noalign{\smallskip}
 Id.  &  z & Main Spectral \\
      & (spectra) & features  \\
\noalign{\smallskip}
\hline\noalign{\smallskip}
A1  &  1.6912 & MgII(2798\AA), FeII(1608\AA), AlII(1670\AA)\\
A2  &  1.6912 & e. line 7518\AA\ \\
B2  &  1.50/2.08  &  e. line 4776\AA \\
B3  &  1.50/2.08  &  e. line 4776\AA \\
C3  &  2.84544 & Ly$\alpha$, CIV~1549\AA \\
E1  &  3.34695 & Ly$\alpha$ \\
S2  &  1.86710& blue cont., abs. lines (see text)\\
V1  &  ? &   no clear feature \\
V2  & 1.2143& Im, [OII]3727\AA \\
V3  & 0.706 & e. line 4776\AA \\
V4  & 2.050 &  CIV~1549\AA \\
V5  & 1.0726& S, e. line 7724.4\AA \\
\hline\noalign{\smallskip}
V6  & 0.40953& Im, [OII]3727\AA, Balmer lines \\
V7  & 0.56692& Im, [OII]3727AA\,[OIII]5007\AA,Balmer lines \\
V8  & 0.58? & S, 4000\AA break, faint abs. lines\\
V9  & 0.41205 & Red S, [OII]3727\AA, Balmer lines \\
V10 & 0.80340 & Red S, [OII]3727\AA, Balmer lines  \\
V11 & 0.38048 & Im, [OII]3727\AA,[OIII]5007\AA \\
V12 & 0.72002 & S, [OII]3727\AA,MgII,H,K,\\
V13 & 0.71965& S, [OII]3727\AA \\
V14 & 0.90031& Im, [OII]3727\AA, Balmer lines, MgII \\
V15 & 0.90233& E, H,K \\
V16 & 0.89935 ?& S, [OII]3727\AA \\
V17 & 0.71169 & S, H, K, Balmer lines \\
V18 & 0.8135 ? &Im, [OII]3727\AA \\
\hline\noalign{\smallskip}
\end{tabular}
\end{flushleft}
\caption[]{ \label{tab_z}
Redshift determination for the 23 observed background galaxies.
The horizontal line separates the core sample, as in Table~\ref{tab_phot}.
For objects with fair S/N on the spectral continuum, the
best fit spectral type is given, from E to Im.
}
\end{table}

\begin{figure*}
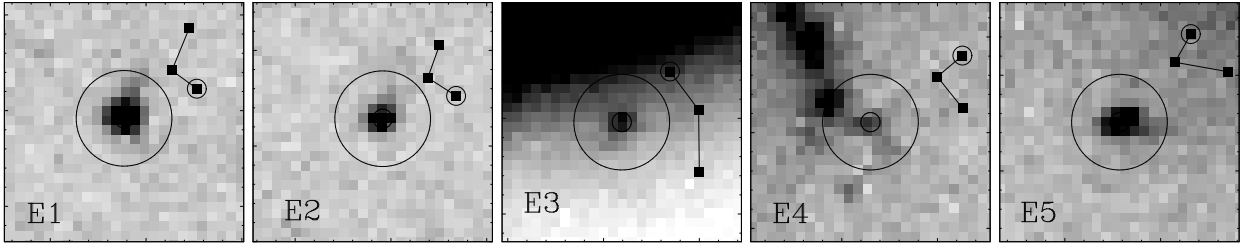

\begin{minipage}{3.2cm}
\psfig{figure=e1.ps,width=3.2cm}
\end{minipage}
\begin{minipage}{3.2cm}
\psfig{figure=e2.ps,width=3.2cm}
\end{minipage}
\begin{minipage}{3.2cm}
\psfig{figure=e3.ps,width=3.2cm}
\end{minipage}
\begin{minipage}{3.2cm}
\psfig{figure=e4.ps,width=3.2cm}
\end{minipage}
\begin{minipage}{3.2cm}
\psfig{figure=e5.ps,width=3.2cm}
\end{minipage}
 
\caption{Morphology of the 5 images of the multiple system E, labelled from E1
  to E5. The size of each image is $2.5\arcsec \times 2.5\arcsec$.  We
  can identify on most images a point source with 2 faint extensions.
  At the top right of each panel, a representation is given to show
  the expected parity of the extensions for all five images. Although
  the image resolution is not good enough to confirm the prediction,
  it seems that the elongations on each images are as expected from
  the lensing theory.}
\label{fig3}
\end{figure*}

The object designated E1 in Table~\ref{tab_z}, showing a strong emission
line with $z=3.347$ when identified with Ly$\alpha$ (see Fig.~\ref{fig2}),
provided a surprising prediction of the NKSE lens model, which to a large
extent was confirmed by the available data. E1, was selected as
a high-z galaxy based solely on its photometric properties; it did not
have a $z_{lens}$ estimate because it appeared as an unresolved
source. The NKSE mass model predicts that an object at the position of
E1 with a redshift of 3.347 is one of the images of a 5 multiple-image lensing
configuration. {\it A posteriori}, a close inspection of the {\it WFPC2} images
reveals E1 as a compact unresolved component with a faint NW
extension. The NKSE model predicted also the positions of the other
four images: a close inspection of the {\it WFPC2} data showed
the predicted images very close to the expected locations(see Fig. \ref{fig1}),
all of them with a morphology similar to E1 (Figure \ref{fig3}) and possibly following
the parity predicted by the lens model. Using E1, E2 and E5, the three
images which are not contaminated by bright objects, we could verify,
within the errors, that they actually have the same spectral energy
distribution(SED), which is
a necessary condition for gravitational images of the same source.

The discovery of the E-system, 
is a very impressive success of the NKSE mass model for AC114,
but at the same time offers the possibility to improve the
model. This is done in the next section.

\begin{table*}
\caption[]{ \label{tab_model}
Characteristics of the new revised model for the mass distribution in AC114.
For each mass component, the following parametres are given: position, ellipticity,
core radius (r$_c$), velocity dispersion and truncation radius (r$_t$). The
galaxy-scale component is given for L$_B^*$.
}
 
\begin{flushleft}
\begin{tabular}{lccccccc}
 
Mass Component & x & y & a/b & $\theta$ & r$_c$ & $\sigma$ & r$_t$ \\
               & (arcsec) & (arcsec) & & (deg) & (kpc) & (km/s) & (kpc) \\
\hline
Central Clump & 0.0$\pm 0.5$ & 0.0$\pm 0.5$ & 2.1 0$\pm 0.05$ & 14.5 $\pm 0.5$
& 97 $\pm 5$ & 1036 $\pm 5$ & 2000 $\pm 200$ \\
Clump 1 & -105.0$\pm 10$ & 0.0$\pm 10$ & 1.25 0$\pm 0.1$ & 40 $\pm 10$
& 100 $(fixed)$ & 620 $\pm 20$ & 1000 $(fixed)$ \\
Clump 2 & 80.0$\pm 5$ & -35.0$\pm 10$ & 1.35 0$\pm 0.1$ & -30 $\pm 10$
& 100 $(fixed)$ & 450 $\pm 20$ & 1000 $(fixed)$ \\
L$_B^*$ Galaxy halo & --- & --- & ---  & ---
& 0.15  & 205$\pm 5$ & 25$\pm 5$\\
 
\end{tabular}
\end{flushleft}
\end{table*}

\section{The lensing mass model revisited}

The NKSE lensing model has been revised in order to include the strong
constraint imposed by the redshifts of two multiple-image systems:
the 3-image S ($z=1.86$), that had been already included in the
NKSE model, and the newly discovered 5-image E ($z=3.347$). 

We follow the NKSE assumptions in tuning up the mass model, and we
refer to their work for more details. In summary, the mass distribution 
is well represented by a cluster-scale component, the central clump, and 
by an additional bimodal cluster-scale component, with the clumps centered 
on the main galaxy concentrations. A galaxy-scale component is also 
introduced, with galaxy halos centered on each bright cluster galaxy,
modelled by a pseudo-isothermal elliptical mass distribution, 
with parameters scaled to the galaxy luminosity (truncation and core radius,
velocity dispersion). To improve the optimisation of
the {\it LENSTOOL} software (Kneib 1993), we used a new
technique involving both a Monte-Carlo search as well as a parabolic
minimization (Golse, Kneib, Soucail 2001).  Only the positions of the
multiple images were fitted, and the free model parameters were the
ellipticity, orientation, velocity dispersion, core radius of the
cluter clump, and the ellipticity, orientation, velocity dispersion
and truncation radius of the central cD galaxy. Table~\ref{tab_model} 
summarizes the characteristics of the new revised model for the 
mass distribution. The best reduced $\chi^2$ found is $ \sim 2$.

Using the best fitted mass distribution, we checked that the mean measured
differences in amplification agree with the ones observed. In
particular for the E-system the measured values, $\Delta m_{E1-E2} = 0.1$ mags
and $\Delta m_{E2-E5} = 0.7$ mags, are in good agreement with model
predictions.

We also used the revised NKSE model to check the redshift prediction
for the other multiple images in the cluster core. We found that the best
redshift estimates for the multiple-image systems A, B, C and D are:
$z_A = 1.7 \pm 0.05$,
$z_B = 1.3 \pm 0.1$,
$z_C = 2.3 \pm 0.1$ and
$z_D = 1.4 \pm 0.1$. These values are larger than the ones previously
predicted by NKSE (see Table \ref{tab_zphot}), resulting in a better
agreement between the $z_{lens}$ and the spectroscopic redshifts (see
Sect. 6).

Table \ref{tab_zphot} lists the lensing redshifts for the multiple-image systems
observed with VLT in the cluster core, and the magnification prediction
for the 23 background galaxies observed with FORS1, using 
the revised model of AC114 and the measured spectroscopic redshifts.

\section{Lensed galaxies in the cluster core}

The identification of spectral features and the main issues concerning
multiple images and highly amplified galaxies are discussed in this
section. 

\subsection{The multiple image A at $z$=1.691}

\begin{figure}
\psfig{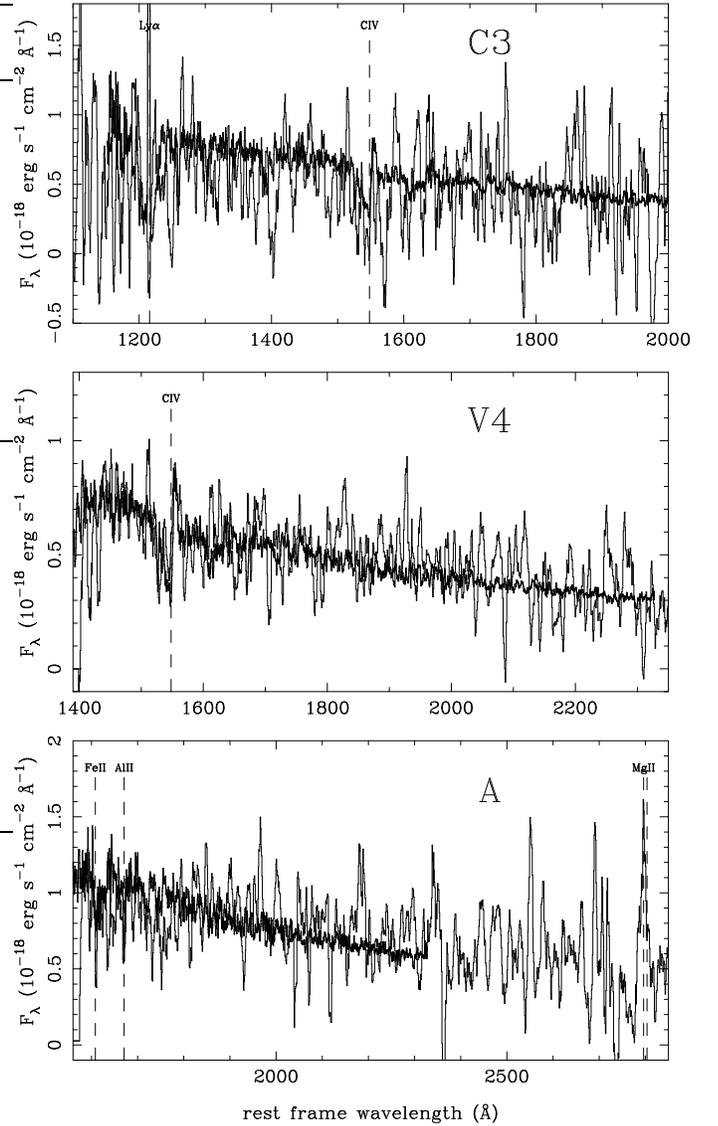}
\caption{From top to bottom, comparison between the spectra of C3 ($z$=2.854), V4 ($z$=2.050)
and A1 ($z$=1.691) (thin lines), and the local starburst galaxy NGC4214 (Leitherer et al 1996)
(thick lines). The main spectral features are shown. 
}
\label{fig5}
\end{figure}

In this case, the redshift is mainly based on a single emission line, which
is found in both the spectra of A1 and A2 at 7518\AA. This line is
identified as MgII (at 2798\AA), in good agreement with the shape of
the continuum.  Besides, the spectrum of A1 also displays absorption lines,
mainly FeII(1608\AA) and AlII(1670\AA). A comparison between the spectra 
of A1 and the local starburst NGC4214 (Leitherer et al 1996) is given in Fig.~\ref{fig5}.
In addition, this redshift value is fully compatible with both the
multiple-image configuration ($z = 1.7 \pm 0.05$ derived from the
revised lensing model) and the $z_{phot}$ (see Table \ref{tab_zphot}). With
$z$=1.691, we do not expect to find other strong emission lines, such
as [OII]3727\AA, CIV~1549\AA\ or Ly$\alpha$, because they are outside 
the observed spectral range (Fig.  \ref{fig5}). 

\subsection{The multiple image S at $z$=1.867}

\begin{figure}
\psfig{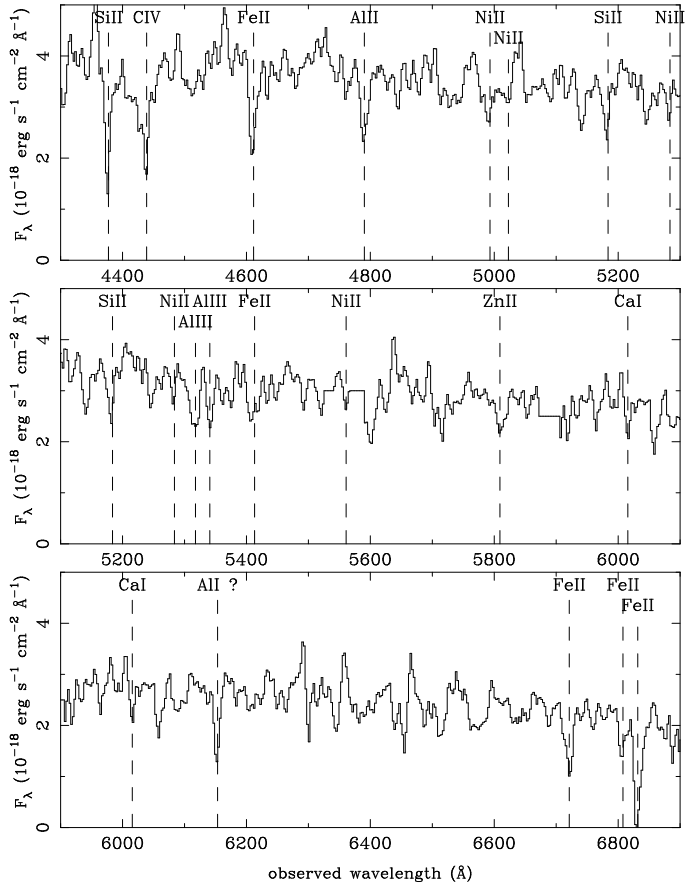}
\caption{Spectrum of object S2, at $z$=1.867, which confirms the
  identification of Smail et al 1995. Note the blue excess of this
  galaxy. }
\label{fig4}
\end{figure}

The previous determination of the redshift of the system S1/S2 by
Smail et al (1995) is confirmed by the present results. We have
obtained a spectrum for S2.  Figure \ref{fig4} displays the high S/N
spectrum as well as the complete line identification. A detailed study 
on this particular object will be presented in a further paper.
This object is also important because it allows a (positive) test of the 
$z_{phot}$ techniques at $1 \le z \le 2$, a redshift domain which is 
particularly difficult to check because of the lack of strong 
spectral features.

\subsection{A new multiple image system (E) at $z$=3.347}

\begin{figure}
\psfig{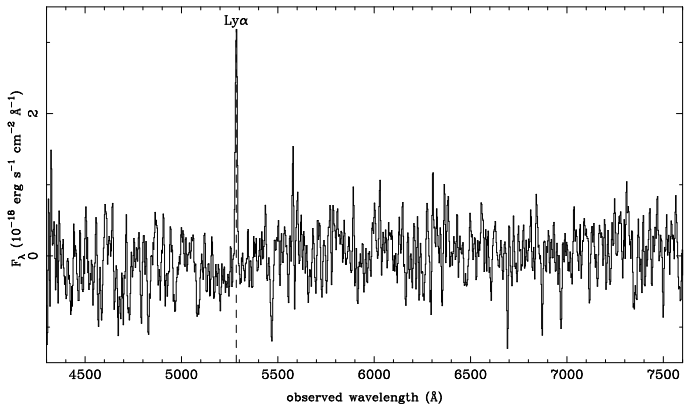}
\caption{Spectrum of object E1, at $z$=3.347, showing a strong emission line, 
  identified as Ly$\alpha$. This redshift implies a 5-image system
  which is actually verified. No other features are clearly identified 
on the continuum. }
\label{fig2}
\end{figure}

\begin{figure}
\psfig{figure=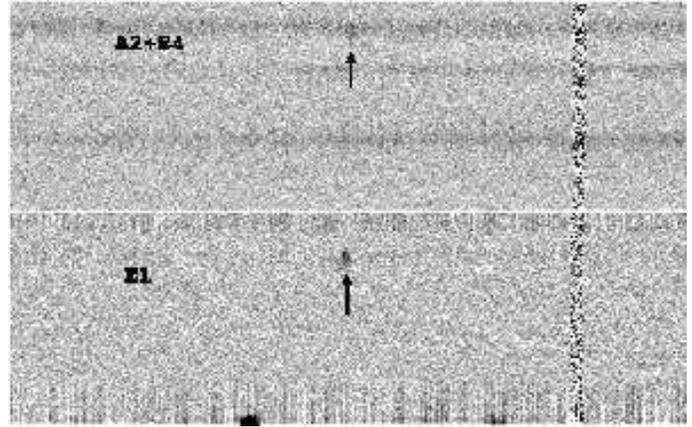,width=0.5\textwidth}
\caption{
  Two dimensional spectra of two different images of the galaxy E.  The arrows indicate the
  position of Ly$\alpha$ line.  The spectrum of the image E4 is blended
  with that of image A2 of arc A, responsible for the spectral continuum
  seen. Note the slight shift along the slit between continuum
  spectrum of A2 and the emission line of E4, which is compatible with their
  position and orientation on the HST image.  }
\label{fig_d26}
\end{figure}

The initial redshift of the system was based on the spectrum taken for
E1. The spectrum, shown in Fig.~\ref{fig2}, shows a strong emission line which
implies $z$=3.347 for the system when it is identified a Ly$\alpha$
line at 1216\AA. As already mentioned, the use of this redshift and
the mass model for AC114 predicted the additional 4 images of the
system which are actually present with essentially the same morphology and 
SED. In fact, this system is one of the two new strong constraints of
the new mass model presented in Sect. 5. Figure \ref{fig_d26} shows the 2D spectra
of E1 and A2+E4; E4 shows a very faint emission line at exactly the same
wavelength as in E1, which is fully compatible with the lensing
interpretation. The equivalent (rest frame) width of Ly$\alpha$ is
$W_{\lambda} = 61 \pm 7 $ \AA. The error bar takes into account the 
uncertainties on the continuum level. 
The corrected absolute magnitude of the source galaxy,
$M_B =-20.6$, together with the observed morphology, suggests a
Seyfert-like galaxy.

\subsection{The multiple images B and C}

\begin{figure}
\psfig{figure=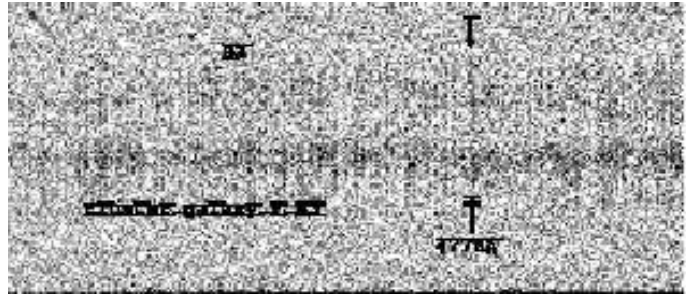,angle=270,width=0.5\textwidth}
\caption{ 
  Rwo dimensional spectra of two different images of the arc B: B2 (top) and B3+B4 (bottom). 
The arrows point the position of a faint emission line, visible in both spectra of 
B2 and B3+B4 images. The B3+B4 spectrum is contaminated by light from a 
nearby cluster galaxy. The slit was placed on the alignment of the B2 image 
with B3 and B4 (see Fig.~\ref{fig1}).}
\label{d21}
\end{figure}

We have obtained spectra for B2 and B3. A single emission line
was found in both spectra, at $\lambda \sim 4776$\AA.  Photometric
redshift could be obtained only for B2 and B1 only, because B3 and B4 are contaminated
by the neighbouring cluster galaxy on most of the ground-based images. The
lensing and photometric redshifts for the image B2, are $z=1.3\pm0.1$
and $z$=$1.61^{+0.19}_{-0.13}$, respectively, at $1\sigma$. 
The $z_{phot}$ for the image B1 is $z$=$1.57^{+0.26}_{-0.25}$,
in excellent agreement with B2. The only plausible
identification for the emission line satisfying these two conditions
is CIII]1909\AA, implying a redshift of $z=1.502$.
However, it does not fit exactly with the lensing prediction
suggesting that either the line identification is not correct, 
or the lens model in this region is not accurate enough.
The emission line is seen on the spectra of both B2 and B3+B4, as
can be appreciated in the 2D spectra shown in Fig.~\ref{d21},
thus it is likely to be a real feature. When it is identified with 
CIV~1549\AA, which is a more likely case, then $z=2.08$, a value 
which is still compatible with the $z_{phot}$ at $2\sigma$. Unfortunately,
in this case Ly$\alpha$ is outside the spectral range observed.
If the lens model in this region 
is responsible for the discrepancy, and the actual redshift of B is 
higher than $z \sim 1.4$, then the redshift prediction for
the system C, $z=2.3\pm0.1$, is also likely to be underestimated.  
Deeper spectroscopy in the I-z bands may allow 
to detect the [OII] emision line, specially if the lensing prediction
is correct.

Our spectrum of image C3 provides a fair estimate of
the redshift of the multiple image C, which shows a blue continuum
with absorption features identified as Ly$\alpha$ and CIV~1549\AA.
The redshift derived from these two lines is $z$=2.854,
which is fully compatible with the $z_{phot}$ estimates.
Figure \ref{fig5} displays a comparison between the spectra of C3 
and that of a local starburst.
Such a spectroscopic redshift is however much larger than the 
present lens prediction,  but in concordance
with the results obtained for the multiple image B.
Clearly better data on B and C are required before deriving any strong
conclusion.

\subsection{The amplified galaxies V1, V2, V3, V4 and V5}

Four other, high redshift, single-image galaxies were observed with
FORS-1:

\begin{itemize}
  
\item V1 is the only source for which we could not identify a clear
  spectral feature, despite a long exposure time (2h47m) using two
  different slit configurations.This object is faint and has a blue
SED. According to its $z_{phot}$ determination, $z$=1.13,
but no strong emission line was detected, in spite of the fact that 
[OII]3727\AA is expected at $\lambda \sim 8000$ \AA. 
  
\item V2 displays a single emission line, superimposed on a continuum
  spectrum, which leads to a redshift $z$=1.2143 when identified with
  [OII]3727\AA.  The $z_{phot}$ determination is in good
  agreement with this spectroscopic redshift.
  
\item The only clear spectral feature in the spectra of V3 is an
  emission line at $\lambda$ = 4776\AA. The $z_{phot}$
  determination gives a 1$\sigma$ interval ranging between 0.81 and
  0.96, with a best fit at $z$=0.84. The most likely identification 
of this  emission line is MgII(2798\AA), leading to $z$= 0.706, a
value which is still compatible with the $z_{phot}$ determination
at $3 \sigma$. Unfortunately, the [OII]3727 \AA~line is expected to 
lie upon the atmospheric [OI] line at 6300 \AA, thus it is hardly 
detected. 
  
\item The spectrum of V4 shows an emission line, with P-Cygni profile,
and a blue continuum.
According to its $z_{phot}$ determination, it is either a high-$z$ source,
with $z$=2.04 (1$\sigma$ error bar ranging between $z$=2.0 and 
$z$=2.07), or a $z$=0.29 foreground galaxy (or even a cluster member
within the errors). In the first case, the emission line is
identified to CIV~1549\AA, leading to $z$= 2.050. In the low-$z$ case,
the emission line is most likely identified with [OII]3727 \AA,
and in this case $z$=0.272, leading to $M_B$ =-17.19.
The latter identification is difficult to
reconcile with the morphology of this object, as it appears on 
the HST images. In addition, the comparison between this spectrum and that
of a local starburst allows to identify other significant absorption features,
which reinforces the high-$z$ identification (Fig.~\ref{fig5}). 

\item V5 displays a single emission line, at $\lambda$ = 7724.4 \AA,
and a relatively blue featureless continuum. According to its $z_{phot}$ determination,
the emission line can be identified with [OII]3727 \AA, leading to 
$z$=1.0726.

\end{itemize}


\begin{table*}
\begin{flushleft}
\begin{tabular}{cccccccc}
\hline\noalign{\smallskip}
 Id.  &  $z$ & $z$ & $\Delta z$ & $z$  & $\Delta m$
 & $M_B$  & Comments \\
      & (spectra)&(phot.)&(phot.)& (lensing) & mag. &  &   \\
\noalign{\smallskip}
\hline\noalign{\smallskip}
\multicolumn{1}{c}{
\begin{tabular}{c}
A1      \\
A2     \\
\end{tabular}
 }   \}    &  1.6912  & 2.04  & 1.27-2.80 & 1.67$\pm$0.15  &  1.58(A1) & $-$20.5 &  \\
\multicolumn{1}{c}{
\begin{tabular}{c}
B2     \\
B3    \\
\end{tabular}
  }   \}   &  1.50/2.08   & 1.61  & 1.48-1.80 & 1.17$\pm$0.10  & 1.85(B2) &  $-$19.5 & $M_
B$ with z=1.50 \\
C3         &  2.854   & 2.76  & 2.67-2.91 & 2.1$\pm$0.3 & 2.40& $-$20.8  & \\
E1         & 3.34695 & 3.45  & 3.29-3.60 & --  & 1.40 & $-$20.6 & \\
S2         & 1.86710 & 2.06  & 1.97-2.25 & --  & 2.18 & $-$21.2 & \\
V1         &   ?       & 1.13   & 1.05-1.22 &   & 1.43 & $-$19.9 & $M_B$ based on $z_{phot}$ \\
V2      &   1.2143   & 0.945  & 0.41-1.08 &  & 0.80 &  $-$20.9 & \\
V3      &   0.706    & 0.84  & 0.81-0.96 &  & 0.48 &  $-$19.1 &  \\
V4      &   2.050    & 0.29/2.04 & 0.26-0.31/2.0-2.07 &  & 0.85 & $-$21.4 &  \\
V5      &   1.0726   & 1.05  & 0.91-1.47 &  & 0.86  & $-$20.7 &     \\
\hline\noalign{\smallskip}
V6      &   0.40953 &  0.46  &  0.38-0.50 &    &  0.142  & $-$19.5  &  \\
V7      &   0.56692 & 0.54  &  0.53-0.55 &    &  0.239   &  $-$21.7 &  \\
V8      &   0.58?  & 0.48   &  0.12-2.08 &    & 0.328    & $-$20.2  &  \\
V9      &   0.41205 &  -     &    -     &     & 0.052  & $-$20.8  &  \\
V10     &  0.80340 &  -    &     -    &       & 0.321  & $-$21.7  &  \\
V11     & 0.38048 &   -     &   -     &       & 0.037  &  $-$19.6 &  \\
V12     & 0.72002 &  0.51   &0.373-0.59 &     & 0.146  & ($-$22.6)&  \\
V13     & 0.71965 &  0.82   & 0.348-2.3 &     & 0.071  & ($-$20.7) &  \\
V14     &  0.90031  & 1.26  &   0.54-3.29   &        & 0.082 &  ($-$20.7) &  \\
V15     &  0.90233 &   -    &   -      &      & 0.074  &  $-$22.7 &  \\
V16     &  0.89935 ? &  -    &   -      &     & 0.067  & $-$22.2  &  \\
V17     &  0.71169 &    -    &    -      &    & 0.040  &  $-$20.6 &  \\
V18     &  0.8135 ?&   -    &    -      &     & 0.053  & ($-$23.9?) &  \\
\hline\noalign{\smallskip}
\end{tabular}
\end{flushleft}
\caption[]{\label{tab_zphot} Characteristics of the background galaxies
studied in AC114. Identifications are the same as in Table~\ref{tab_phot}.
$M_B$ have been corrected for magnification $\Delta m$ according to the
revised NKSE model. The $z_{lens}$ values are from NKSE(1998).
Error bars in $z_{phot}$ correspond to $1 \sigma$ using also
a B-image. $M_B$ given
in brackets correspond to objects for which the restframe B band has been
extrapolated from the available photometry, using the best fit model. It is
highly model dependent in the case of V18.}
\end{table*}
 
\subsection{Absolute Magnitudes in the spectroscopic sample}

The revised mass model has been used to derive the
magnification of each of the images.  The absolute magnitudes M$_B$
for these amplified galaxies are computed through a direct scaling of
the observed SED, taking into account the spectroscopic $z$, and using
the best-fit templates from the Bruzual \& Charlot code (Bruzual \&
Charlot 1993) to derive the k-corrections. The M$_B$ values, corrected
for amplification according to the revised NKSE model,
are given in Table \ref{tab_zphot}.
The absolute magnitudes of these galaxies typically range from M$_B$
$\sim -19$ to $-22$, with a median value of $-$20.5. In the cluster core,
where the amplification is the highest, absolute magnitudes range between
M$_B$ $\sim -19$ and $-21.5$. In particular, E and C, the two objects 
with confirmed redshifts in the interval
$2.5 \ltapprox z \ltapprox 3.5 $, are respectively 0.5 
and 1.5 magnitudes fainter than the limiting magnitude of the Steidel 
et al (1999) sample at similar redshifts. In the case of S, the magnification 
corrected magnitude is close to the limiting value in conventional samples, 
but the magnification of $\sim 2$ magnitudes per image allows the
obtention of spectra with higher signal-to-noise ratios.

\subsection{Redshift distribution of lensed galaxies in the photometric sample}

\begin{figure}
\psfig{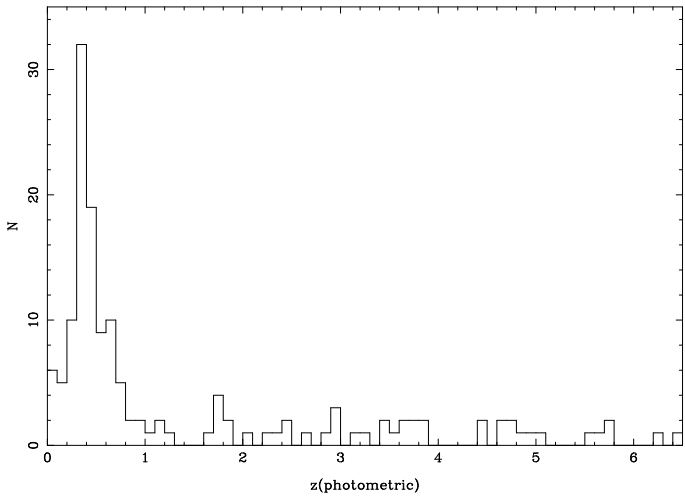}
\caption{Raw photometric redshift distribution of 
galaxies in the $1.2\arcmin \times 1.2\arcmin$ cluster core.
\label{histo_z}
}
\end{figure}

One of the main goals of this program is to recover the intrinsic properties
of a very faint subsample of high-$z$ lensed galaxies, undetectable otherwise, using 
deep images and $z_{phot}$, in order to investigate, in particular,
their $z$ distribution and luminosity 
function. We have obtained through {\it hyperz} the N$(z_{phot})$
distribution of arclets in the $1.2\arcmin \times 1.2\arcmin$ core of AC114,
a region covered by all the images in Table~\ref{tab1}. Fig. \ref{histo_z}
displays the raw N$(z_{phot})$ distribution for the 148 objects considered in 
this region. This sample includes objects detected in at least 3 
filters. For $z_{phot}$ calculations, undetected objects in a given band have 
their flux in this band set to zero, 
with a flux error corresponding to the local limiting magnitude.
Photometric errors are taken from SExtractor,
with a threshold zeropoint error of 0.1 magnitudes.
>From this sample, we exclude all cluster-member
candidates on the basis of their SEDs. Eight foreground and 81 background galaxies remain
in the final sample, defined as objects excluded as cluster members to better
than 90\% confidence level, with $z \le 0.2$ or $z \ge 0.4$ . Forty-four galaxies are 
found at $1 \le z \le 7 $ in a $1.2\arcmin \times 1.2\arcmin$ field centered on the cD galaxy.
For comparison, the total number of such sources expected in this field, taking 
into account the depth of the survey, ranges from $\sim 30$ to 50 sources.
The selection criteria allow that the background sample includes all the objects 
selected for spectroscopy in the cluster core. 

\begin{figure}
\psfig{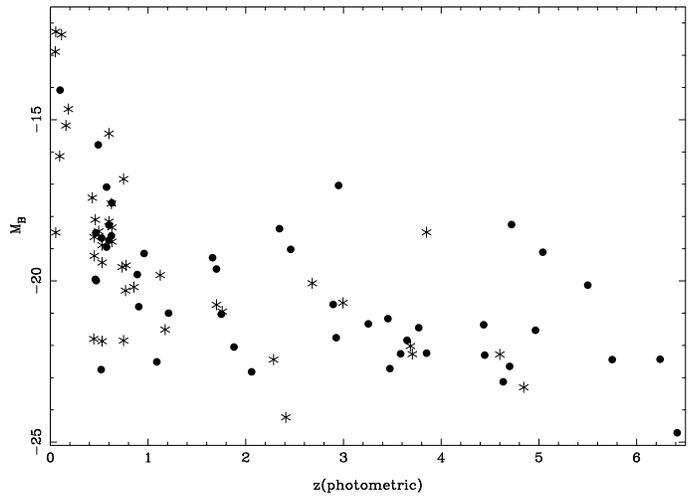}
\caption{Lens corrected absolute magnitude distribution versus redshift
for sources in the $1.2\arcmin \times 1.2\arcmin$ cluster core. Black dots 
and stars correspond to objects with reduced $\chi^2 \le 1.5$ and
$\chi^2 > 1.5$, respectively.
\label{MB_z}
}
\end{figure}

For cluster lenses with well constrained mass distributions, as in the present case, 
it is possible to recover precisely the $N(z_{phot})$ and the absolute magnitude distributions 
of lensed galaxies by correcting the relative impact parameter on each redshift bin. 
We have used the NKSE corrected mass model to derive such distributions for the background 
sample of galaxies. Fig. \ref{MB_z} displays the lens corrected absolute magnitude M$_B$
distribution versus redshift for the sources in the field of AC114.
Up to $z \sim 4$, the restframe 
B band is well matched by the filters used here. The median value of M$_B$ is -21.5 for 
the 44 galaxies at $1 \ltapprox z \ltapprox 7 $, with typical values ranging between 
-19 and -23. In the $1 \ltapprox z \ltapprox 3.5 $ redshift interval (23 galaxies), the 
median M$_B$ is -21.0. Thus, the spectroscopic sample is biased towards the most amplified
sources in the central region, and thus intrinsically fainter than the photometric sample.
The median intrinsic luminosity increases to M$_B \sim -22$ for the $z \ge 3.5$ sample.
Because of the small number of objects in the sample,
this exercise is of limited interest for a single cluster, when using standard 
(not ultra-deep) imaging, but it is useful to illustrate the case.

\section{Discussion}

This study has allowed to improve the NKSE lensing model for
the massive and well studied cluster AC114, mainly through the
discovery of a new multiple-image system, the E-system. This discovery
was driven by the identification of a very high-z galaxy in the core
of AC114 thanks to deep multi-band images,
and specifically through the photometric redshift technique. This
multiple-image system was missed by a search of lensed objects in the
high-resolution HST images due to its point-like appearance.

An indication of the precision that can be attained for the
intrinsic luminosities of the background galaxies is obtained from the
independent inferences, based on the revised mass model, that can be
made from the photometry of the individual images belonging to a
given multiple image system. The typical uncertainty in the magnification 
factor is $\sim 0.3$ magnitudes for most objects, and this gives an 
idea of the limitations arising from lens modelling when using 
gravitational telescopes. 

The iterative optimization of the
mass model together with feedback between the lens and photometric
redshifts in this stage, makes highly desirable to have deep
multi-band image. Thus, high-resolution images in addition to 
deep multi-band images are actually needed for the selection of high-z galaxy
candidates and the obtation of sufficient constraints to establish a
robust mass model. This could be done for a master list of clusters
that could be made public through archives. Once the mass model is
mature for a cluster, such as for AC114 now, deeper and higher resolution
images, and occasional spectroscopy, should conduct to key information
and new discoveries in the distant universe.

The galaxy cluster AC114, with its highly constrained
gravitational-lens model, is a particularly powerful gravitational
telescope for which further lensing studies can be undertaken, such as
searching for distant supernovae (Sullivan et al 2000), 
distant Submm sources (Smail et
al 1997) or Lyman-$\alpha$ emitters (Ellis et al, in preparation).
By performing a very accurate mass modelling we can also expect to constraint
the cosmological parameters ($\Omega$, $\lambda$) using
the method of Golse et al (2001).  

The source responsible for the multiple image E at $z=3.347$ is intrinsically
faint, roughly the equivalent of the local $M^*_B + 0.5$ magnitudes.
The presence of a relatively strong emission line, together with the 
gravitational magnification, has allowed us to obtain a spectroscopic 
redshift. This galaxy is intrinsically $\sim 0.5$ magnitudes fainter 
than the limiting value of the Steidel et al (1999) sample at similar 
redshifts. It has a bright core and at least two faint extensions, similar to
the morphology of the compact cores in the field sample of Steidel et al
(1996a) at $2.5 \ltapprox z \ltapprox 3.5$. 
The main subclump displays a compact morphology, which is not
resolved in width on the WFPC2 images, thus implying length scales of the 
order of $\sim 1 kpc$. Its morphology and the equivalent (rest frame) 
width of Ly$\alpha$ are similar to those of H5, one of the multiple
images at z=4.05 identified in the core of A2390 (Pell\'o et al 1999),
and also similar to the typical values observed by Hu et al (1998)
in their sample of emission line galaxies at $z \sim 3$ to 6.

Six galaxies in our spectroscopic sample are found at $1 \ltapprox z
\ltapprox 2.5$, a redshift domain which is relatively poorly known 
because of the lack of strong spectral features allowing to identify 
sources in this redshift interval. The gravitational amplification 
has allowed to identify such objects, and the
precise redshift determination , using visible and near-IR
spectrographs in 8-10m class telescopes, will allow a detailed
study of the spectral energy distributions of
these $z\ge$1 galaxies. In particular, studies 
of the star formation rate history and of the 
permitted region in the age-metallicity-reddening parameter space, are
now possible and will be presented in the future. The best example is
the source S, for which a detailed spectroscopic study is presently 
in progress.

The median value of M$_B$ is -21.5 for the photometric sample of 
galaxies at $1 \ltapprox z \ltapprox 7 $, with typical values ranging between 
-19 and -23. The median M$_B$ is -21.0 in this sample, in the  
$1 \ltapprox z \ltapprox 3.5 $ redshift interval, and it increases to
M$_B \sim -22$ for the $z \ge 3.5$ sample. The spectroscopic sample 
in the cluster core is biased towards the most amplified sources, 
which are intrinsically fainter than the photometric sample, and it
exhibits a median M$_B$ of -20.5 in the
$1 \ltapprox z \ltapprox 3.5 $ redshift bin, with absolute magnitudes ranging
between M$_B$ $\sim -19$ and $-21.5$. Lensed galaxies in this spectroscopic sample 
are found to be intrinsically fainter, between 0.5 and 1.5 magnitudes, 
than the limiting magnitudes
in the present {\it blank field} surveys at $2 \ltapprox z \ltapprox 
3.5$ (Steidel et al 1999).
This gain in sensitivity towards low luminosity high-z objects,
equivalent to an increase in collecting area by at least a factor of 2 
to 3 (and up to a factor of $\sim 10$ in restricted regions of the source plane), 
is sufficiently large to allow a substantial improvement on our knowledge 
of the faintest and/or highest redshift galaxies.

\section{Conclusions}
 
In summary, we have successfully identified 23 galaxies in the background of
AC114, 10 of them being high-redshift objects, with $0.7 \ltapprox z
\ltapprox 3.5$, located on the cluster core. The main results of our
spectroscopic survey, concerning 
multiple-image systems and the mass model of this cluster, are:

\begin{itemize}

\item We confirm the redshift $z$=1.87 of the gravitational pair S1/S2.

\item The redshift of the multiple image system A has been measured: $z$=1.69,
in excellent agreement with the lensing and photometric predictions.

\item A new 5-image multiple system E at redshift $z$=3.347 is identified.
We use this new strong constraint, together with the previously used 
multiple system S, to derive an improved mass model for AC114.

\item The redshifts of the multiple images B and C are determined, however
new and better quality data are needed to confirm them. So far, their
current determinations
are slightly larger than those predicted by the original NKSE, and still
difficult to reconcile with the revised model.

\end{itemize}

Overall, these results show the
validity of the approach and the good accuracy of the predictions
of the lensing model,
and clearly establish with great confidence a mass model for AC114:
the revised NKSE model.

We have demonstrated the efficiency of a lensing/photometric
approach to derive new samples of high-$z$ galaxies, conditioned to
the availability of high-resolution images and a fairly well
determined mass model for the cluster-lens. All the highly 
magnified galaxies in the spectroscopic sample have 
M$_B$ $\sim -21.5$ to $-19$ opening a way to study the
faint end slope of the luminosity function in early epochs of the universe.
In particular, the 2 objects (C,E) with spectroscopic $2.5 \ltapprox z 
\ltapprox 3.5 $, are respectively 0.5 and 1.5 magnitudes fainter 
than the limiting magnitude of the Steidel et al (1999) survey  
at similar redshifts. 
We plan to continue efforts to gather a sample of very faint 
high-$z$ lensed galaxies employing the
$z_{phot}$ technique described in this work, and to investigate
their spectral energy distributions using visible and near-IR spectrographs
in 8 or 10-m class telescopes. A galaxy sample
of the proper size to study their luminosity
function will require observations and analyses,
similar to the one presented here for AC114,
of about ten massive cluster-lenses.

\acknowledgements

We are grateful to M. Bolzonella, G. Bruzual, M. Dantel-Fort, G. 
Mathez and D. Schaerer for useful discussions on this particular program. We would
like to thank Dr. W. J. Couch for providing his deep B image of AC114,
and Dr. C. Leitherer for allowing the use of the NGC 4214 spectrum. LEC was
partially supported by FONDECYT grant 1970735. 
Part of this work was supported by the French {\it Centre National de la
  Recherche Scientifique}, by the French {\it Programme National de
  Cosmologie} (PNC), and the TMR {\it Lensnet} ERBFMRXCT97-0172
(http://www.ast.cam.ac.uk/IoA/lensnet) and the ECOS SUD Program.
Based on observations collected at the European Southern Observatory,
Chile (ESO N° 64.O-0439), and with the NASA/ESA Hubble Space
Telescope, which is operated by STScI for the Association of
Universities for Research in Astronomy, Inc., under NASA contract
NAS5-26555.

\end{document}